\pdfoutput=1
\documentclass[11pt]{article}

\usepackage[T1]{fontenc}
\usepackage[utf8]{inputenc}
\usepackage{lmodern}
\usepackage{geometry}
\usepackage{array}
\usepackage{longtable}
\usepackage{booktabs}
\usepackage{url}
\usepackage[hidelinks]{hyperref}
\usepackage{amssymb}

\newcolumntype{L}[1]{>{\raggedright\arraybackslash}p{#1}}

\title{Conversational AI for Automated Patient Questionnaire Completion: Development Insights and Design Principles}
\author{%
David Fraile Navarro\textsuperscript{1} and Mor Peleg\textsuperscript{2}\\[0.5em]
\textsuperscript{1}Macquarie University, Sydney, Australia\\
\textsuperscript{2}University of Haifa, Haifa, Israel
}
\date{}

\begin{document}
\maketitle

\begin{abstract}
Collecting patient-reported outcome measures (PROMs) is essential for clinical care and research, yet traditional form-based approaches are often tedious for patients and burdensome for clinicians. We developed a generative AI conversational agent(CA) using GPT-5 to collect back pain data according to the NIH Task Force's Recommended Minimal Dataset. Unlike prior CAs that ask questions one-by-one, our CA engages users in topic-based conversations, allowing multiple data items to be captured in a single exchange. Through iterative development and pilot testing with clinicians and a consumer panel, we identified key design principles for health data collection CAs. These principles extend established clinical decision support design guidelines to conversational interfaces, addressing: flexibility of interaction style, personality calibration, data quality assurance through confidence visualization, patient safety constraints, and interoperability requirements. We present our prompt design methodology and discuss challenges encountered, including managing conversation length, handling ambiguous responses, and adapting to LLM version changes. Our design principles provide a practical framework for developers creating conversational agents for patient questionnaire completion. The CA is available at \url{https://chatgpt.com/g/g-68f4869548f48191af0544f110ee91c6-backpain-data-collection-assistant} (requires ChatGPT registration and subscription for unlimited use).
\end{abstract}


\section{Introduction}
Reliable collection of data from back-pain patients is crucial for assessment of their diagnosis, prognosis, progress tracking, and for prospective studies that assess the effectiveness of therapies. The National Institute of Health Task Force on Research Standards for Chronic Low Back Pain defined a standard minimal dataset [1] that should be collected from patients to enable comparison studies on patient populations and improve patient care. The dataset includes over 40 data items and scales for reporting each item. It is very tedious for patients to report a complete and accurate dataset. Elicitation of this data from patients by automated tools could help save time of medical personnel and relieve their burden of documentation, which is known to be a serious problem for medical personnel [2][3][4].

We developed a customized CA that aims to easily and accurately collect data from patients who have experienced back pain. The CA guides the user in articulating responses that each cover a certain topic that spans several questions at a time (e.g., symptoms, impact on daily functioning, impact on sleep); a task that is not trivial for humans to do.

The emergence of large language models (LLMs) has created transformative opportunities for automating patient data collection. A recent landmark randomized controlled trial with over 2,000 patients demonstrated that an LLM-based CA reduced specialist consultation duration by 28.7\% while substantially improving care coordination [5]. At scale, conversational AI has been deployed to collect standardized mental health assessments from over 64,000 patients across multiple clinical services [6]. These advances have prompted researchers to envision a new generation of "LLM-PROMs": patient-reported outcome measures that combine individualized conversational items with real-time AI interpretation, potentially overcoming the limitations of traditional structured questionnaire formats [7].

In this paper we focus on the insights that we gathered during the CA's development and testing process and formulate them as design principles. Our design principles validate and extend related research that presented design principles for clinical decision support systems and conversational agents for collecting patient-reported outcome measures (PROMs).

\section{Related work}
Other researchers have developed CAs for collecting PROMs and have reported their design principles. Chen et al. [8] introduced design principles used to design Chat-ePRO. For each item in the PROMs questionnaire they generated a section in the prompt containing the item's question and options. Based on the conversation, the CA constructed an output dataset in that format. In addition, the prompt instructed the CA to refrain from altering the content of the question when asking. Furthermore, to ensure patient safety, the conversations were reviewed by clinicians so that in case of potential harm patients could be contacted.

Comparative studies have demonstrated measurable advantages of conversational interfaces over traditional forms. Soni et al. [9] directly compared a CA versus an online form for health data collection, finding that the CA achieved higher System Usability Scale scores (69.7 vs 67.7) and significantly higher Net Promoter Scores (24 vs 13), indicating greater user satisfaction. Similarly, Nguyen et al. [10] reported substantially higher usability scores for CA-based family health history collection compared to form-based approaches (SUS 80.2 vs 61.9), with personalization ranked as the highest priority enhancement by participants. However, Wilczewski et al. [11] found that while users had favorable experiences with shorter CA interactions, satisfaction declined with longer conversations, a critical consideration when designing CAs for comprehensive questionnaires like the NIH back pain dataset.

Maharjan et al. [12] introduced Can We Talk -- a speech-enabled conversational agent for questionnaire-driven self-report of health and wellbeing. The authors recommend employing conversational cues to support engagement. Specifically, they recommend tailoring the CA's response in real-time to the emotional valence of users' comments, striking an empathetic tone following a negative user response. They also suggest keeping questions simple and providing the meaning of complex terms in order to avoid requiring users to guess the meaning of terms and response options.

Wei et al. [13] attempted to use prompt engineering based on GPT-3 to collect user self-reported information, unrelated to health. The study analyzed the impact of different prompts on generation of questions and summarized potential errors that may arise. Based on their experience, the authors recommend: 1) using structured format for the questions and answers, 2) specifying a personality modifier for the CA (e.g., empathetic) so that the CA could generate appropriate acknowledgment messages; 3) instructing the CA to follow different questioning paths based on user responses to specific questions to avoid asking questions that are not necessary (e.g., if the user did not undergo an operation, additional questions about this topic need not be asked); 4) collected information in multi-stage prompts for long data collection, because they noted that with the conversation goes longer, CAs have a tendency to miss information slots that appear later in prompts.

Understanding user acceptance of health CAs requires attention to broader trust factors. The Technology Acceptance Model (TAM) provides a foundational framework for health technology adoption; a seminal review demonstrated that TAM and its extensions can explain up to 70\% of variance in behavioral intention [14]. Design principles specific to health conversational agents have been articulated in systematic reviews. Kocaballi et al. [15] found that personalized feedback and adaptive responses improve engagement and dialogue quality in healthcare CAs, though theoretical frameworks for personalization remain underdeveloped. Fadhil and Schiavo [16] proposed core principles including building empathy through emotional sensitivity, ensuring CA personality guides users appropriately, and acknowledging that no one-size-fits-all approach exists for health CA design. The most comprehensive paper that we found that focused on design principles was not specific to CAs but for clinical decision-support systems (CDSS), in general. Horsky et al. [17] focused on principles for designing CDSS that are usable and make sure that they are clinically relevant, accurate, specific, and clear. Six of their eight principles are also relevant for CAs for health form collection; they are presented in Capital letters in Table 1.

\section{Design principles and recommendation features}
We adopted the design principles presented by Horsky et al. [17] that were relevant to health CAs. We removed two design principles that were out of scope for health form completion: Advice: assessment, suggestion and recommendation -- not commands and Maintenance and re-use of intermediate variables. We added design features specific to CAs, based on the related work and our own experience, that go beyond Horsky et al. They are presented in italics in Table 1.

\begin{longtable}{L{0.30\textwidth}L{0.62\textwidth}}
\caption{Summary of desired system attributes.}\label{tab:table1}\\
\toprule
\textbf{Design principle} & \textbf{Design feature recommendations for form-completion tasks}\\
\midrule
\endfirsthead
\toprule
\textbf{Design principle} & \textbf{Design feature recommendations for form-completion tasks}\\
\midrule
\endhead
\bottomrule
\endfoot
\textbf{Consistency} of design concepts, visual formats, and terminology &
-Provide clear response options: \textit{number options for long option sets, especially when they involve parentheses and commas, so that the LLM will not lose track}\\
& \textit{-Customize the standard questionnaire options, when essential, to your local setting (e.g., terms related to ethnicity and race)}\\
& - Keep questions simple and providing the meaning of complex terms [12]\\
& - Use meaningful color sets \textit{to convey additional information, such as tiered certainty or severity levels (see Assure Data quality)}\\
\midrule
\textbf{Flexibility of interaction and conversation style} to enhance user experience &
\textit{-LLMs are a good choice for supporting flexibility as it is inherent in them, yet prompts should be designed to restrict some of that flexibility}\\
& \textit{-Converse rather than strict question by question form completion, with the following features: a) short dialogs; b) free conversation – not presenting questions as technical as in the survey; c) conversation chunked into discussion topics so the user could address several questions at once; d) follow different questioning paths based on user responses to specific questions [13]; e) ability to adapt the conversation style in order to communicate in the user's preferred way.}\\
& \textit{-Users should be able to switch between the three input methods: form, chat, or voice}\\
\midrule
\textbf{\textit{CA personality}}\textit{ [13] modified to be appropriate} &
\textit{-Instruct and test the personality of the AI assistant to be fun but appropriate (no slang and respectable), displaying empathy [12] in the "right amount"}\\
\midrule
\textit{Assure }\textbf{\textit{data quality}} &
\textit{-Communicate to the user the data captured so that the user could confirm or revise, and clearly communicate the degree of certainty that the data was captured correctly, for example using traffic-light colors}\\
& \textit{-Check that the CA collects all data; if questions appearing toward the end of long conversations produce poor data collection, separate into multi-stage prompts [13]}\\
\midrule
\textit{Maintain patient's compliance to complete the reporting [8] by }\textbf{\textit{providing encouragement}} &
\textit{-Occasionally let the user know how much more is needed}\\
& \textit{-Motivate the user to complete the questionnaire}\\
\midrule
Presentation of information in a way that \textbf{cultivates trust} &
\textit{-Evidence-based questionnaire}\\
& \textit{-Control on dialog (tone, no advice – see Flexibility of Interaction)}\\
\midrule
\textit{Patient }\textbf{\textit{safety}} &
\textit{-Limit the conversation to data collection based on the evidence-based recommended data set without offering medical advice}\\
\midrule
Periodic review of system inferences and human actions (\textbf{logs}) &
Allow access to logs, analyze periodically to increase the \textit{quality of data collection and allow tracing the data collected to the relevant conversation}\\
\midrule
Interoperability and \textbf{data standards}, integrity and robust architecture &
Normalize data collected to a common \textit{structured} representational format (instructions in prompt) [8][13]. Later, the data could be translated to standards such as FHIR, OMOP.\\
\midrule
Innovation and third-party developers &
\textit{-Use LLM environments of certified vendors (e.g., openAI, Open WebUI)}\\
& \textit{-Fix the version of the LLM to a version that you have tested so that the experience for user would be controlled. Occasionally test a newer version of the LLM.}\\
& -Allow certified companies access to data services, interface development; separate code and content\\
\end{longtable}

\section{Design methods}
In this section. we follow the Chatbot Assessment Reporting Tool (CHART) [18] for reporting the details methods that we (first author DFN, MD, PhD, and second author MP, Professor of Information Systems) used to implement the recommendations and design features (presented in Section 3). These include the prompt [19] used to build the CA. Our detailed methods are organized into six subsections, following the order of implementation.

\subsection{Basing data collection on an evidence-based patient-reported outcome questionnaire}
In our case it was based on the Report of the NIH Task Force on research standards for chronic low back pain [1].

\subsection{Structure and content of the prompt for the CA}
Table 2 provides the design principles for the prompts. The full prompt is provided in [19]. We developed the prompt iteratively; the first iteration included elicitation of just 5 data items according to the Recommended Minimal Dataset [1] and was done using the OpenAI playground platform. It focused on defining and checking the output coded data that was collected and on the tone/personality of the CA. The subsequent six iterations were done as a custom GPT (see Section 4.6). Given that the CA generates text with a certain extent of randomness (temperature parameter), the iteration focused on providing instructions to reduce the randomness and define the required interaction, the wording of questions and answers, explanations provided of the user, conveying certainty of collected data, and provision of feedback and encouragement (see second column of Table 2). We also drew ideas introduced by the CA in some (but not all) of the interactions we had with it, and fixed them in the prompt. The main idea that we drew from the CA was to use color to display the tiered degree of confidence that the CA had in the data it has collected from the user. Iterations 1-7 were done with GPT-4o. Iterations 8-9 were done with the OpenWebAI platform, using GPT 5.0 version of August 7, 2025.

\begin{longtable}{L{0.30\textwidth}L{0.62\textwidth}}
\caption{Sample of Prompt's design principles and the text from the actual prompt.}\label{tab:table2}\\
\toprule
Design principle & Text from the actual prompt\\
\midrule
\endfirsthead
\toprule
Design principle & Text from the actual prompt\\
\midrule
\endhead
\bottomrule
\endfoot
\begin{minipage}[t]{\linewidth}
-CA's Role (Medical interviewer)\newline
-Specialty of the CA (collecting data)\newline
-Description of the storage format for the "Recommended Minimal Dataset"
\end{minipage}
&
\begin{minipage}[t]{\linewidth}
You are a very knowledgeable medical interviewer,\newline
…specializing in collecting data about back pain.\newline
The data that needs to be collected has to be stored using the data-items corresponding to the questions specified in the "Recommended Minimal Dataset", for each question -one answer provided by the patient user. The stored data should be recorded in the following "Storage Format":\newline
\texttt{<Question number>: <Answer number>}\newline
Eg Q1:A3; Q2:A2; Q3:A2...
\end{minipage}\\
\midrule
Description of how the CA needs to start the conversation and the goal (good user experience) &
For a good user experience, please start by asking the user to introduce themselves and their back pain. Based on the Recommended Minimal Dataset questions, you could provide a short explanation of what they are expected to describe.\\
\midrule
\begin{minipage}[t]{\linewidth}
\textbf{Design principles} for conversation\newline
-Short dialog\newline
-Free conversation\newline
-Discussion of topics: several questions at once\newline
-Traffic light reflects what the CA captured from their dialog\newline
-Reflect (feedback) to the user the CA's level of confidence in each data\newline
-For patient safety\newline
-For retaining user engagement\newline
-CA's tone empathetic yet professional
\end{minipage}
&
\begin{minipage}[t]{\linewidth}
This is how you need to do it:\newline
-short dialogs\newline
-Free conversation; don't present questions as technical as in the survey.\newline
-Break into discussion topics so the user could address several questions at once\newline
-After you collected responses for a topic, provide a short summary of the data. For example,"\\checkmark{} Q14:A3 — Off work/unemployed due to back pain: Does not apply".\newline
Use green color when your confidence is high, red when you have no clue, and amber for cases in between.\newline
-don't offer treatment advice\newline
- Occasionally let the user know how much more is needed.\newline
- Make it fun but appropriate for your role (no slang and respectable), empathy in the right amount.
\end{minipage}\\
\midrule
Description of the minimal data set &
The following "Recommended Minimal Dataset" has 41 data items that need to be collected from back pain patients. It is presented as multiple-choice questions for which only one answer should be selected. The possible answers are presented below the question, in parentheses, separated by commas.\\
\midrule
The first data item to be collected (out of 41) &
Q1. How long has low-back pain been an ongoing problem for you? \textbf{If there was pain years ago (eg, injury at childhood), which was resolved and later relapsed, please answer about the current episode.}\newline
(Less than 1 month, 1–3 months, 3–6 months, 6 months–1 year, 1–5 years, > 5 years)\\
\midrule
Questions broken into topics with a leading sentence before each topic &
Question 5 has 4 parts (a-d) related to symptoms and how they have bothered them. For each, the answer is (Not bothered at all, Bothered a little, Bothered a lot)\\
\midrule
Question dependencies &
Only if you patient replied "yes" to question 6, please ask questions 7 and 8\\
\midrule
Local Setting-specific questions modified &
Questions 35 and 36 will be skipped because related to race options relevant to USA\\
\midrule
For long option sets, the options are numbered A1,A2, etc. &
37. Employment Status\newline
A1-Working now A2-Looking for work A3-unemployed A4-Sick leave or maternity leave A5-Disabled due to back pain: permanently or temporarily A6-Disabled for reasons other than back pain A7-Student A8-Temporarily laid off A9-Retired A10-Keeping house A11-Other\\
\midrule
Providing details for "Other" &
If you selected other, please specify\\
\end{longtable}

All iterations used temperature of 0.8 and a context window 128,000; total tokens 16,384; presence penalty 0; frequency penalty: 0. Iteration 8 was prototyped by two researchers from Macquarie University's Center for Health Informatics. Following their evaluation, we revised the prompt to contain instructions regarding the units of measure of height and weight, the numbering of multiple-choice options for long option sets, and revised the prompt such that it would not contain repetitions of the same instructions. Later evaluations by clinicians and a patient consumer panel did not necessitate prompt revision. The CA was accessed from Australia and Israel.

\subsection{Increasing patient safety}
The prompt for the CA directs it to not provide any therapy recommendations, limiting the risk to patient safety. A potential, yet minimal risk could be related to data collection. ChatGPT (and is underlying AI models respond to user's queries for further explanations and might proactively add explanations to the questions they pose that were not in the original back pain questionnaire. To limit the risk that the CA may interpret the questions of the back pain questionnaire incorrectly, we inspected the questionnaire [1] to see where there could be any questions that could be unclear to participants. We added clarifications to the text of the questions by consulting a medical doctor. We supplied this extended questionnaire to the CA. Our inspection yielded only one question (Q1) that needed clarification; the clarifying text is shown in bold in Table 2, Row 6.

Some questions contained everyday terms in the English language that could need clarifications for some users. We therefore analyzed the 40 conversations by the 4 users who participated in our pilot study. We found one question where the CA proactively supplied further explanations to user; it supplemented "exercise therapy" with correct examples: "physio, stretching programs, etc.". We also found terms in four questions for which users could need further explanation and asked the CA for clarification. These were: \#12:chores, \#15:worker's compensation, \#19:run errands, \#26:problems with sleep (See [19] for full questionnaire and prompt). We checked several times and the CA was consistent and its answers were correct and did not result in any misinterpretation. These safety measures were found as sufficient by two clinicians from and eight members of a consumer forum from Macquarie University and by the University of Haifa Ethics Committee (approval \#472/25).

\subsection{Data quality assurance}
During our pilot with the CA, we found that there were times when the CA accepted user's input without employing common-sense quality assurance. For example, when a user was asked to enter height (in cm) and weight (in Kg) but entered them in the opposite order (59, 165), the CA didn't ask for clarification. Yes, in other cases, the CA did employ reasoning to complete missing information. For example, when the user was asked about 4 aspects of sleep quality and said that she had some problems falling asleep, the CA inferred "somewhat" problems in falling asleep and also somewhat problems in sleep. The CA's acceptance of erroneous information regarding height and weight may reflect the known shortcomings with interpreting numbers. This could be alleviated with specific code for quality assurance.

\subsection{Instructions for users to enhance their experience}
In addition to explaining to users about their safety and the focus on data collection rather than therapy recommendations, we also instructed users that if they feel that they want the tone of the CA to be different or would like it to ask more or fewer questions at a time, or ask it to clarify the questions, they could do so.

\subsection{Choosing a CA platform}
We experimented with three platforms and decided to use the third.

1) We used OpenAI playground platform (\url{https://platform.openai.com/playground/prompts}) to test prompts for the CA

2) Then we created a custom GPT (\url{https://chatgpt.com/gpts}) using a prompt, which we could share with users and include in an app. However, users have to have an OpenAI account, and to use tokens, which cost money, in order to interact with the CA for more than a few sentences. Therefore, we chose a third platform:

3) OpenWebUI - a self-hosted, open-source AI platform that provides a user-friendly chat interface for interacting with large language models (LLMs). We used it to interface to OpenAI. By entering our OpenAI API key, we could allow users to use our tokens so that they could complete their conversations without needing to expose their real emails and real names and the usage payment is from our account. We created a web platform to handle the experimental set up as well as connecting to our self-hosted OpenWebUI server in Amazon Web Services (AWS). We also fixed the version of the underlying LLM (gpt-5-2025-08-07) so no further updates will affect CA's integrity. We purchased a domain name: backpainbot.academy from which users could interface with the OpenWebUI and the experimental platform.

\section{Preliminary Evaluation}
We evaluated the CA with a) eight members of the Australian Institute of Health Innovation (AIHAI) Consumer Engagement Panel, which supports a person-centered approach to research across all stages of the research lifecycle. Six of them were patients, one was a clinician, and one was a patient advocate; b) with a physiotherapist specializing in pain management; and c) with a student and a staff member from Macquarie University's Center for Health Informatics.

The evaluation with the consumer panel and with the physiotherapist each lasted about 30-40 minutes in a videoconference. We presented a short video demo [20] of the CA conversing with a back pain patient and asked the participants for their comments. Specifically, we asked the consumers whether the CA could add value relative to form-filling and whether the CA's responses to the user in the demo were clear, relevant, and that the tone of the CA was appropriate. We asked the physiotherapist another question, whether she saw benefit in asking patients to interact with the CA while they are waiting for their appointment with their care provider; the benefit could be for obtaining a high-quality data set and for saving clinician's time (less documentation).

The pain specialist had very positive impressions of the CA's usability and its usefulness in eliciting pain data from patients. She suggested that it could be extended to support collection in a single dialog of multiple forms that are collected in practice in Australia for pain management. She also emphasized that in order to promote patients' trust, it is important to explain to patient who the creators of the CA are, and that the CA was instructed to collect their data based on evidence-based standard back pain questionnaires. The consumer panel also had positive impressions about the potential added value of the CA; they thought that the CA was engaging and valued the clarifications it provided and the fact that it helped the patient reflect on the information recorded by the CA. However, two consumers noted that the CA was too verbose and its sentences were too long; users could be instructed to optionally ask the CA to change its interaction according to their needs and a user guide about logging into the system could support less competent users. Consumers asked whether the CA could support conversation in multiple languages and whether it could be used in voice mode (it can but we have only evaluated its support for English via text interface). They also asked about privacy and emphasized that the information collected by the CA should not be abused, and that each patient should have access the full information that was gathered from him.

\section{Discussion and Conclusions}
Our iterative development process revealed several important insights about LLM behavior in health data collection contexts that informed the design principles presented in Table 1.

\textbf{Emergent behaviors requiring control.} Without explicit prompting guardrails, the CA exhibited behaviors that, while well-intentioned, were inappropriate for a medical data collection context. Most notably, the CA tended to be overly friendly, enthusiastic, or colloquial, adopting a tone inconsistent with how a healthcare professional would address back pain issues. Similarly, the CA sometimes displayed excessive empathy that could be perceived as forced or inauthentic, particularly since users were aware they were interacting with an AI rather than a human. This observation aligns with recent findings that while empathetic CA responses enhance perceived warmth, they may paradoxically reduce perceived authenticity [21]. These observations led to our design principle emphasizing that CA personality should be "fun but appropriate," with empathy calibrated to the "right amount."

\textbf{Spontaneous helpful behaviors.} Interestingly, some beneficial behaviors emerged without explicit instruction. In one early conversation, the CA spontaneously introduced traffic-light color coding to indicate its confidence level in captured data, a feature we had not specified. Recognizing its value for data quality assurance, we formalized this behavior in subsequent prompt iterations. This illustrates how iterative testing with generative AI can surface useful design features that developers may not have anticipated.

\textbf{Domain-specific constraints.} Two areas required explicit constraints in the prompt. First, the CA needed clear instructions to avoid offering medical advice, limiting its role strictly to data collection. Second, certain questions in the NIH dataset required clarification that the CA could not reliably provide without guidance, particularly regarding the temporal and episodic nature of chronic back pain and flare-ups. We addressed this by adding clarifying text directly to the prompt for ambiguous questions.

\textbf{Practical implications.} Table 1 can serve as a design checklist for developers creating health data collection CA. Our experience suggests that while LLMs offer remarkable flexibility for conversational interfaces, this flexibility must be carefully constrained through prompt engineering to ensure appropriate professional tone, calibrated empathy, and domain-appropriate behavior.

\textbf{Multimodal considerations.} Although we did not implement it for logistical constraints with the experimental set up, we envisaged that patients could switch between the three input methods (form, chat, voice): this is also what patient told us in the panel could be helpful for certain populations. Specific prompts are required to handle different modalities (e.g. limit verbosity, adapt for oral speech patterns of communication). Further developments may include a multiple-questionnaire filling CA (where data is collected for several health questionnaires simultaneously) that was requested from healthcare professionals, as they confirmed how cumbersome is both for patients and professionals to complete all the mandatory questionnaire filling required with health care interactions, especially for chronic patients.

\subsection{Limitations}
Proprietary LLMs showed different behavior depending on the version and with updates, when the same prompt was supplied to them. We noticed that prompts that generated the expected output for ChatGPT 4o did not work as expected with GPT-5 or required further tuning. We also noticed that the personality of these models is severely affected by its training/fine-tuning/reinforcement-learning (e.g., for a given prompt 4o was "friendly enough" while 5.1 was "too friendly"). Developers should exercise caution and re-test when the CA is updated to a newer model, or if a "silent update" (not communicated to the user) is suspected. For these reasons we recommend sticking to one model for the whole development cycle and if possible fixed to a given, identifiable time-point (e.g. gpt-4o-20250831 or similar) and accessed through API endpoints.

\subsection{Future work}
Future work should explore voice interfaces, which require different prompt designs to convey collected data and confidence levels without visual cues like traffic-light colors. Additionally, CAs could be extended to gather contextual information beyond the standard questionnaire, such as detailed pain history, which emerged as valuable during our pilot conversations. There's also the potential for this CAs to continue gathering data over time, or store previous medical records, such as with recent developments like ChatGPT Health [22] which could further enhanced the role of these in the healthcare domain and for reporting comprehensive PROMs.

\subsection{Conclusions}
We developed a generative AI CA for collecting back pain data according to the NIH Task Force's Recommended Minimal Dataset and documented the design principles that emerged from our iterative development process. Our experience demonstrates that while LLMs provide powerful capabilities for conversational health data collection, they require careful prompt engineering to maintain appropriate professional tone, calibrate empathetic responses, ensure patient safety, and handle domain-specific ambiguities. The design principles presented in Table 1, which extend established CDSS usability guidelines to conversational interfaces, provide a practical framework for developers creating similar health data collection CAs. Our preliminary evaluation with clinicians and a consumer panel yielded positive feedback, suggesting that dialogue-driven approaches to patient questionnaire completion merit further investigation in clinical settings.

\section*{Acknowledgements}
David Fraile Navarro is supported by NHMRC grant GNT2008645. We thank Kalissa Brooke-Cowden and Dr. Priyanka Rana from Macquarie University's Center for Health Informatics, Dr. Cynthia Ashley and the Australian Institute of Health Innovation Consumer Panel for their participation in the pilot evaluation and their useful comments.


\begin{thebibliography}{99}
\bibitem{ref1}
Deyo RA, Dworkin SF, Amtmann D, Andersson G, Borenstein D, Carragee E, et al.
Report of the NIH Task Force on research standards for chronic low back pain. Phys. Ther. 2015;95(2):e1-8.

\bibitem{ref2}
Levy DR, Withall JB, Mishuris RG, Tiase V, Diamond C, Douthit B, et al.
Defining documentation burden (DocBurden) and excessive DocBurden for all health professionals: a scoping review. Appl. Clin. Inform. 2024;15(5):898--913.

\bibitem{ref3}
Holmgren AJ, Adler-Milstein J, Apathy NC.
Electronic Health Record Documentation Burden Crowds Out Health Information Exchange Use By Primary Care Physicians: Article examines electrnoic health record documentation burden. Health Aff. 2024;43(11):1538--45.

\bibitem{ref4}
Moy AJ, Schwartz JM, Chen R, Sadri S, Lucas E, Cato KD, et al.
Measurement of clinical documentation burden among physicians and nurses using electronic health records: a scoping review. J. Am. Med. Informatics Assoc. 2021;28(5):998--1008.

\bibitem{ref5}
Tao X, Zhou S, Ding K, Li S, Li Y, Wu B, et al.
An LLM chatbot to facilitate primary-to-specialist care transitions: a randomized controlled trial. Nat. Med. 2026;\url{https://doi.org/10.1038/s41591-025-04176-7}.

\bibitem{ref6}
Rollwage M, Habich J, Jueche K, Carrington B, Viswanathan S, Stylianou M, et al.
Using Conversational AI to Facilitate Mental Health Assessments and Improve Clinical Efficiency Within Psychotherapy Services: Real-World Observational Study. JMIR AI. 2023;(2):e44358.

\bibitem{ref7}
Terheyden JH, Pielka M, Schneider T, Holz FG, Sifa R.
A new generation of patient-reported outcome measures with large language models. J. Patient-Reported Outcomes. 2025;9(1):34.

\bibitem{ref8}
Chen Z, Wang Q, Sun Y, Cai H, Lu X.
Chat-ePRO: Development and pilot study of an electronic patient-reported outcomes system based on ChatGPT. J. Biomed. Inform. 2024;154:104651.

\bibitem{ref9}
Soni H, Ivanova J, Wilczewski H, Bailey A, Ong T, Narma A, et al.
Virtual conversational agents versus online forms: patient experience and preferences for health data collection. Front. Digit. Heal. 2022;(4):954069.

\bibitem{ref10}
Nguyen MH, Sedoc J, Taylor CO.
Usability, engagement, and report usefulness of Chatbot-based family health history data collection: Mixed methods analysis. J. Med. Internet Res. 2024;(26):e55164.

\bibitem{ref11}
Wilczewski H, Soni H, Ivanova J, Ong T, Barrera JF, Bunnell BE, et al.
Older adults' experience with virtual conversational agents for health data collection. Front. Digit. Heal. 2023;(5):1125926.

\bibitem{ref12}
Maharjan R, Rohani DA, Bardram JE, Doherty K.
Can we talk ? Design Implications for the Questionnaire-Driven Self-Report of Health and Wellbeing via Conversational Agent. In: Proceedings of the 3rd Conference on Conversational User Interfaces. 2021. p. 1--11.

\bibitem{ref13}
Kim JWS, Jung H, Kim YH.
Leraging large language models to power chatbots for collecting user self-reported data. In: Proceedings of the ACM on Human-Computer Interaction. 2024. p. 1--35.

\bibitem{ref14}
Holden RJ, Karsh BT.
The technology acceptance model: its past and its future in health care. J. Biomed. Inform. 2010;43(1):159--72.

\bibitem{ref15}
Kocaballi AB, Berkovsky S, Quiroz JC, Laranjo L, Tong HL, Rezazadegan D, et al.
The personalization of conversational agents in health care: systematic review. J. Med. Internet Res. 2019;21(11):e15360.

\bibitem{ref16}
Fadhi A, Schiavo G.
Designing for health chatbots [Internet]. 2019;Available from: arxiv:1902.09022

\bibitem{ref17}
Horsky J, Schiff GD, Johnston D, Mercincavage L, Bell D, Middleton B.
Interface design principles for usable decision support: a targeted review of best practices for clinical prescribing interventions. J. Biomed. Inform. 2012;45(6):1202--16.

\bibitem{ref18}
Huo B, Collins GS, Chartash D, Thirunavukarasu AJ, Flanagin A, Iorio A, et al.
Reporting guideline for Chatbot Health Advice studies: the CHART statement. JAMA Netw. open. 2025;8(8):e2530220-.

\bibitem{ref19}
Peleg M, Navarro DF.
Promt for backain data collection Conversational Agent [Internet]. 2025;Available from: \url{https://docs.google.com/document/d/19CspcDiNDkGl4WbR58SLilEfIpDFC3Y7CSrlTHH-4a8/edit?usp=sharing}

\bibitem{ref20}
Peleg M, Navvaro DF.
Backpain ChatBot Video Demo [Internet]. 2025;Available from: \url{https://youtu.be/074xFsD9jGY}

\bibitem{ref21}
Seitz L.
Artificial empathy in healthcare chatbots: Does it feel authentic? Comput. Hum. Behav. Artif. Humans. 2024;2(1):100067.

\bibitem{ref22}
OpenAI.
Introducing ChatGPT Health [Internet]. 2026;Available from: \url{https://openai.com/index/introducing-chatgpt-health/}
\end{thebibliography}
\end{document}